\documentstyle[prl,aps,multicol,twoside]{revtex}

\begin{document}

\draft

\title{Universal correlations of one-dimensional interacting electrons
in the gas phase}

\author{F. G\"ohmann and V. E. Korepin}

\address{Institute for Theoretical Physics, State University of New
York at Stony Brook, Stony Brook, NY 11794-3840, USA}

\maketitle

\begin{abstract}
We consider dynamical correlation functions of short range interacting
electrons in one dimension at finite temperature. Below a critical
value of the chemical potential there is no Fermi surface anymore, and
the system can no longer be described as a Luttinger liquid. Its low
temperature thermodynamics is that of an ideal gas. We identify the
impenetrable electron gas model as a universal model for the gas phase
and present exact and explicit expressions for the asymptotics of
correlation functions at small temperatures, in the presence of a
magnetic field.
\end{abstract}

\pacs{PACS: 71.27.+a, 71.10.Ca, 71.10.Fd, 71.10.Pm}

\begin{multicols}{2}

\thispagestyle{empty}

There are two different phases of a one-dimensional system of
interacting electrons. In order to explain the difference, let us first
consider zero temperature, $T = 0$:

At zero temperature there exists a critical value $\mu_c$ of the
chemical potential. If $\mu > \mu_c$, the density of electrons in the
ground state is positive. This is the familiar Luttinger liquid phase
\cite{Haldane}. Correlations are dominated by fluctuations around the
Fermi surface. Conformal field theory \cite{BPZ84} describes their
power law decay.

If $\mu < \mu_c$, the density at $T = 0$ is zero. This is a special
case of the electron gas phase, which will become less trivial at
positive temperatures.

What happens at small positive temperatures, $T > 0$?

For $\mu > \mu_c$ the Luttinger liquid phase persists. Now correlations
decay exponentially. In order to describe correlations at small
temperature one has to employ a conformal mapping from the complex
plane to a strip of finite width. As a result the rate of exponential
decay is defined by conformal dimensions.

Similarly, the gas phase persists for $\mu < \mu_c$ and $T > 0$. The
density of electrons is now positive, but typically exponentially
small at small temperatures. The ideal gas law holds. This suggested
the name `gas phase'. Correlations in the gas phase behave essentially
different compared to those in the Luttinger liquid phase. This is the
subject of this paper.

We shall argue below that the low temperature behaviour of correlation
functions in the gas phase is universal for different models. The
universal model which determines the asymptotics of correlation
functions of one-dimensional electrons with short-range repulsive
interaction is the impenetrable electron gas, which is the infinite
coupling limit of the electron gas with repulsive delta interaction.
The universality class of the impenetrable electron gas comprises a
large number of physically interesting systems, most notably the
Hubbard model and its non-solvable generalizations.

The impenetrable electron gas model is solvable by Bethe ansatz
\cite{YangGaudin,Takahashi71b}. Based on the Bethe ansatz solution
a determinant representation for the dynamical two-point Green
functions was obtained in \cite{IzPr}, and, subsequently, their
asymptotics was calculated in \cite{GIKP98,GIK98}. The case of static
correlations was treated earlier in \cite{Berkovich91}.

\section{The Hubbard model in the gas phase}
In order to get a better understanding of the gas phase let us start
with the important example of the Hubbard model,
\begin{eqnarray}
     H_H & = & - \sum_{j=1}^L (c_{j+1, \sigma}^+ c_{j, \sigma} +
			 c_{j, \sigma}^+ c_{j+1, \sigma})
         + U \sum_{j=1}^L n_{j \uparrow} n_{j \downarrow} \nonumber \\
	 && \quad \quad \quad \quad
	    - \mu \sum_{j=1}^L (n_{j \uparrow} + n_{j \downarrow})\ .
\end{eqnarray}
Here the canonical Fermi operators $c_{j, \sigma}^+$, $c_{j, \sigma}$
are creation and annihilation operators of electrons at site $j$
of a one-dimensional periodic lattice of length $L$, and
$n_{j, \uparrow}$ and $n_{j, \downarrow}$ are the corresponding
particle number operators. $U$ is the strength of the Hubbard
repulsion, $\mu$ is the chemical potential.

The Hubbard model was solved \cite{LiWu68} by means of the nested Bethe
ansatz. This allows us to test our ideas about the gas phase
quantitatively. The energy levels for the $N$ electron system are
\begin{equation} \label{energy}
     E = 2 \sum_{j=1}^N (1 - \cos k_j) - (\mu + 2) N\ ,
\end{equation}
where the charge momenta $k_j$ are solutions of the Lieb-Wu equations
\cite{LiWu68}. Clearly the first term on the right hand side of
(\ref{energy}) is non-negative. Hence, if $\mu < \mu_c = - 2$, the
energy of all eigenstates becomes non-negative, and the absolute ground
state is the empty lattice. For $\mu > - 2$, on the other hand, the
energy can be lowered by filling states with small $k$'s. Since
$k_{j+1} - k_j \sim 1/L$, this leads to a finite density of electrons
in the ground state as $L \rightarrow \infty$. We conclude that the
Hubbard model is in the gas phase for $\mu < - 2$ and in the Luttinger
liquid phase else. The asymptotics of correlation functions of the
Hubbard model in the Luttinger liquid phase was obtained in
\cite{FrKo}.

The thermodynamics of the Hubbard model was first considered by
Takahashi \cite{Takahashi72}. He expressed the Gibbs free energy
$\omega = - P$ ($P$ pressure) in terms of the dressed energies $\kappa
(k)$, $\varepsilon_n (\Lambda)$, $\varepsilon_n' (\Lambda)$, of
elementary excitations at finite temperature. $\kappa (k)$ is the
dressed energy of particle (or hole) excitations, $\varepsilon_n
(\Lambda)$ describes spin excitations and $\varepsilon_n' (\Lambda)$
$k$-$\Lambda$ strings. All $k$-$\Lambda$ strings are gapped
\cite{Takahashi72}. They do not contribute to the low temperature
thermodynamic properties of the Hubbard model \cite{Takahashi74} and
drop out of the equation for the pressure, which simplifies to
\begin{equation}
     P = \, \frac{T}{2 \pi} \int_{- \infty}^\infty dk \,
	 \ln \left( 1 + e^{- \frac{\kappa (k)}{T}} \right)\ .
\end{equation}
Similarly, the integral equations for the dressed energies at low
temperatures become
\begin{eqnarray}
     && \kappa (k) = - \mu - 2 \cos k \nonumber \\ &&
		     \quad \quad \quad \quad
		     - \, T \sum_{n=1}^\infty
		     \left( [n] \ln \left( 1 +
			e^{- \frac{\varepsilon_n}{T}} \right) \right)
			(\sin k)\ , \label{kappa} \\
     && \ln \left( 1 + e^{\frac{\varepsilon_n}{T}} \right) = \nonumber
	\\ && \quad \quad \quad
	   - \int_{- \pi}^\pi dk \, \cos k \, a_n (\Lambda - \sin k)
	     \ln \left( 1 + e^{- \frac{\kappa (k)}{T}} \right) \nonumber
	\\ && \quad \quad \quad
	   + \sum_{m=1}^\infty \left( A_{nm}
	     \ln \left( 1 + e^{- \frac{\varepsilon_n}{T}} \right)
	     \right) (\Lambda)\ , \label{epsn}
\end{eqnarray}
where $n = 1, 2, 3, \dots$ in equation (\ref{epsn}), and
\begin{equation}
     a_n (\Lambda) = \frac{nU/4 \pi}{(nU/4)^2 + \Lambda^2}\ .
\end{equation}
$[n]$ and $A_{nm}$ are integral operators defined by
\begin{eqnarray}
     && ([0]f) (\Lambda) = f(\Lambda)\ , \\
     && ([n]f) (\Lambda) = \int_{- \infty}^\infty d \Lambda' \,
	a_n (\Lambda - \Lambda') f(\Lambda')\ , \ n = 1, 2, \dots \\
     && A_{nm} = \!\!\!\!\! \sum_{j=1}^{{\rm min} \{n, m\}}
		 \!\!\!\!\!\!\!
		 \left( \big[|n - m| + 2(j - 1) \big] +
		 \big[|n - m| + 2j \big] \right).
\end{eqnarray}
The gas phase is characterized by the absence of a Fermi surface for
$\kappa (k)$. Thus $\kappa (k)$ is positive in the zero temperature
limit, and the first term on the right hand side of (\ref{epsn}) becomes
exponentially small in $T$. Dropping this term, the equations
(\ref{epsn}) decouple from (\ref{kappa}). Since the equations become
independent of $\Lambda$, it is not hard to solve them. The solution
is the same as in the infinite coupling limit $U \rightarrow \infty$
(cf.\ e.g.\ \cite{Takahashi71b}), $\exp\{ \varepsilon_n (\Lambda)/ T\}
= n(n + 2)$. Inserting this solution into (\ref{kappa}) we obtain
\begin{equation} \label{kappagas}
     \kappa (k) = - \mu - 2 \cos k - T \ln 2\ .
\end{equation}
Our initial assumption that $\lim_{T \rightarrow 0} \kappa (k) > 0$ for
all $k$ is self-consistent, if $\mu + 2 < 0$, which is precisely the
condition for being in the gas phase stated above. With (\ref{kappagas})
the low temperature expression for the pressure becomes
\begin{equation} \label{gibbshu}
     P = \, \frac{T}{2 \pi} \int_{- \infty}^\infty dk \,
      	 \ln \left( 1 + 2 e^{\frac{\mu + 2 \cos k}{T}} \right)
            \approx \, \sqrt{\frac{T}{\pi}} \, e^{\frac{\mu + 2}{T}}\ ,
\end{equation}
and we see that density $D = \partial P/\partial \mu$ and pressure $P$
are related by the ideal gas law,
\begin{equation} \label{ideal}
     P = TD\ .
\end{equation}

There are two important lessons to learn from our simple calculation.
First, the low temperature limit in the gas phase works the same way
as the strong coupling limit at finite temperatures. Second, the low
temperature Gibbs free energy $\omega = - P$ in the gas phase shows no
signature of the lattice. It is the same as for the impenetrable
electron gas (see below), which is a continuum model. This fits well
with our intuitive understanding of the gas phase: The mean free path
of the electrons is large compared to the lattice spacing $\Delta$,
their momentum is small. Their kinetic energy is of the order $T$.
Hence, the effective repulsion is large for $T \ll U$.

\section{Scaling}
The above arguments suggest that an effective Hamiltonian that
describes the Hubbard model in the gas phase can be obtained as the
continuum limit of the Hubbard Hamiltonian. Let us introduce the
continuous length $\ell = \Delta L$ and coordinates $x = \Delta n$
connected with the $n$th lattice site. In the continuum limit, $\Delta
\rightarrow 0$ for fixed $\ell$, we obtain canonical field operators
$\Psi_\sigma (x)$ for electrons of spin $\sigma$ as
\begin{equation} \label{psi}
     \Psi_\sigma (x) = \lim_{\Delta \rightarrow 0}
		       \frac{c_{n, \sigma}}{\sqrt{\Delta}}\ .
\end{equation}
Let us perform the rescaling
\begin{eqnarray}
     && T_H = \Delta^2 T \quad, \quad \mu_H + 2 = \Delta^2 \mu \quad,
	\quad k_H = \Delta k\ , \nonumber \\
     && t_H = t/\Delta^2 \quad, \quad B_H = \Delta^2 B\ , \label{scale}
\end{eqnarray}
where $k$ denotes the momentum, $t$ the time and $B$ the magnetic field,
which we shall incorporate below. The index `$H$' refers to the Hubbard
model. Then, in the limit $\Delta \rightarrow 0$, we find
\begin{equation}
     H_H/T_H = H/T\ .
\end{equation}
Here $H$ is the Hamiltonian for continuous electrons with delta
interaction,
\begin{eqnarray}
     && H = \int_{- \ell/2}^{\ell/2} dx \Big\{
            (\partial_x \Psi_\alpha^+ (x)) \partial_x \Psi_\alpha (x)
	    \nonumber \\ && \quad
	    + \frac{U}{\Delta} \Psi_\uparrow^+ (x) \Psi_\downarrow^+ (x)
	      \Psi_\downarrow (x) \Psi_\uparrow (x)
	    - \, \mu \Psi_\alpha^+ (x) \Psi_\alpha (x)\Big\}.
	    \label{gasham}
\end{eqnarray}
Note that the coupling $c_1 = U/\Delta$ of the continuum model goes to
infinity! This is a peculiarity of the one-dimensional system. The
effective interaction in the low density phase becomes large. Similar
scaling arguments lead to an effective coupling $c_2 = U$ in two
dimensions and to $c_3 = \Delta U$ in three dimensions, i.e.\ unlike
one-dimensional electrons three-dimensional electrons in the gas phase
are free.

\section{Universality}
What happens to more general Hamiltonians in the continuum limit?
Let us consider Hamiltonians of the form $H_G = H_H + V$, where
$H_H$ is the Hubbard Hamiltonian and $V$ contains additional short
range interactions. We shall assume that $V$ is a sum of local
terms $V_j$ which preserve the particle number. Then $V_j$ contains as
many creation as annihilation operators, and the number of field
operators in $V_j$ is even. We shall further assume that $V_j$ is
hermitian and space parity invariant.

According to equation (\ref{psi}) every field $c_{j, \sigma}$ on the
lattice contributes a factor of $\Delta^{1/2}$ in the continuum limit.
One factor of $\Delta$ is absorbed by the volume element $dx = \Delta$,
when turning from summation to integration. Thus, if $V_j$ contains 8
or more fields, then $V \sim \Delta^3$ and $V/T_H$ vanishes. If $V_j$
contains 6 fields, then at least two of the creation operators and two
of the annihilation operators must belong to different lattice sites,
since otherwise $V_j = 0$. A typical term is e.g.\ $V_j =
c_{j, \uparrow}^+ c_{j, \downarrow}^+ c_{j + 1, \uparrow}^+
c_{j + 1, \uparrow} c_{j, \downarrow} c_{j, \uparrow}$. In the continuum
limit we have $c_{j + 1, \uparrow} = \Delta^{1/2} \Psi_\uparrow (x)
+ \Delta^{3/2} \partial_x \Psi_\uparrow (x) + O (\Delta^{5/2})$.
Hence, the leading term vanishes due to the Pauli principle. The next
to leading term acquires an additional pow\-er of $\Delta$. We conclude
that $V \sim \Delta^3$ and thus $V/T_H \rightarrow 0$.

If $V_j$ contains 4 fields, then
\begin{equation} \label{4fields}
     V \sim \Delta^2 \Psi_\uparrow^+ (x) \Psi_\downarrow^+ (x)
	\Psi_\downarrow (x) \Psi_\uparrow (x) + O (\Delta^4)\ .
\end{equation}
Here the first term on the right hand side is the density-density
interaction of the electron gas. In order to arrive at the impenetrable
electron gas model the coefficient in front of this term has to be
positive. Note that there are no terms of the order of $\Delta^3$ on
the right hand side of (\ref{4fields}) and thus no other terms than
the first one in the continuum limit. Terms of the order of $\Delta^3$
would contain precisely one spatial derivative. They are ruled out,
since they would break space parity.

Considering the case, when $V_j$ contains 2 fields, we find, except
for the kinetic energy and the chemical potential term, terms which
correspond to coupling to an external magnetic field $B_H$. For these
terms to be finite in the continuum limit we have to rescale the
magnetic field as $B_H = \Delta^2 B$ (cf.\ equation (\ref{scale})).

Our considerations show that the impenetrable electron gas model with
magnetic field,
\begin{equation} \label{hb}
     H_B = H + B \int_{- \ell/2}^{\ell/2} dx \;
	   \Psi_\alpha^+ (x) \sigma_{\alpha \beta}^z \Psi_\beta (x)\ ,
\end{equation}
is indeed the universal model (for small $T$) for the gas phase of
one-dimensional lattice electrons with repulsive short-range
interaction.

\section{Impenetrable electrons}
The impenetrable electron gas is the infinite coupling limit of the
electron gas with repulsive delta interaction ($\Delta \rightarrow
\infty$ in (\ref{hb})), which was the first model solved by nested
Bethe ansatz \cite{YangGaudin}. The pressure of the system as a
function of $T$, $\mu$ and $B$ is known explicitly \cite{Takahashi71b},
\begin{equation} \label{gibbsif}
     P = \, \frac{T}{2 \pi} \int_{- \infty}^\infty dk \,
	 \ln \left( 1 +  e^{\frac{\mu + B - k^2}{T}}
		             +  e^{\frac{\mu - B - k^2}{T}} \right)\ ,
\end{equation}
and may serve as thermodynamic potential. The expression (\ref{gibbsif})
is formally the same as for a gas of free spinless fermions with
effective (temperature dependent) chemical potential $\mu_{eff} = \mu +
T \ln(2 \cosh B/T)$. Hence the Fermi surface vanishes for
$\lim_{T \rightarrow 0} \mu_{eff} = \mu + |B| < 0$. The finite
temperature correlation functions of the impenetrable electron gas
depend crucially on the sign of $\mu_{eff}$. This allows us to define
the gas phase at finite temperature by the condition $\mu_{eff} < 0$,
which is also sufficient for deriving the ideal gas law (\ref{ideal})
from the low temperature limit of (\ref{gibbsif}). Note that for zero
magnetic field and small temperature equation (\ref{gibbsif}) coincides
with the rhs of (\ref{gibbshu}).

The time and temperature dependent (two-point) Green functions are
defined as
\begin{eqnarray} \label{gp1}
     G_{\uparrow \uparrow}^+ (x,t) & = &
		 \frac{{\rm tr} \left( e^{- H/T} \,
        \Psi_\uparrow (x,t) \Psi_\uparrow^+ (0,0) \right)}
	{{\rm tr} \left( e^{- H/T} \right)}\ , \\ \label{gm1}
     G_{\uparrow \uparrow}^- (x,t) & = &
		 \frac{{\rm tr} \left( e^{- H/T} \,
        \Psi_\uparrow^+ (x,t) \Psi_\uparrow (0,0) \right)}
	{{\rm tr} \left( e^{- H/T} \right)}\ .
\end{eqnarray}
For the impenetrable electron gas these correlation functions were
represented as determinants of Fredholm integral operators in
\cite{IzPr}. The determinant representation provides a powerful tool
to study their properties analytically.

The short distance asymptotics can be obtained `perturbatively' by
expanding the determinant representation for small $x$ and $t$ and
fixed ratio $k_0 = x/2t$,
\begin{eqnarray} \label{gpshort}
     G_{\uparrow \uparrow}^+ (x,t) & = &
	\frac{e^{- \frac{{\rm i} \pi}{4}}}{2 \sqrt{\pi t}}
	+ \frac{e^{\frac{{\rm i} \pi}{4}}}{2 \sqrt{\pi}}
	(k_0^2 + \mu - B) \sqrt{t} \nonumber \\
	&& \; - \left(1 - \frac{1}{\gamma}\right) \frac{D}{\pi}
	   - \left(1 + \frac{1}{\gamma}\right) \frac{D}{2}
	   + O (t)\ , \\ \label{gmshort}
     G_{\uparrow \uparrow}^- (x,t) & = & D_\uparrow
	+ E_\uparrow \, {\rm i} t
	+ O \left(t^{3/2}\right)\ .
\end{eqnarray}
Here $\gamma = 1 + e^{B/T}$, $D_\uparrow = \partial P/
\partial (\mu - B)$ is the density of up-spin electrons and
\begin{equation}
     E_\uparrow = \frac{1}{2 \pi} \int_{- \infty}^\infty \! dk
		  \frac{(k^2 - \mu + B) e^{- \frac{B}{T}}}
		       {2 \cosh(B/T) + e^\frac{k^2 - \mu}{T}}
\end{equation}
may be interpreted as the `energy' of the up-spin electrons. In the gas
phase at low temperatures the density $D$ becomes exponentially small
and the terms proportional to $D$ may be neglected in (\ref{gpshort}).
Then, to lowest order in $D$, (\ref{gpshort}) and (\ref{gmshort}) are
the same as for free fermions.

In \cite{GIKP98,GIK98} the determinant representation was used to
derive a nonlinear classical differential equation, which drives the
correlation functions. This equation is closely related to the quantum
Hamiltonian (\ref{gasham}). It is the separated nonlinear Schr\"odinger
equation. Together with a corresponding Riemann-Hilbert problem it
determines the large-time, long-distance asymptotics of the correlators
(\ref{gp1}), (\ref{gm1}). In \cite{GIKP98,GIK98} the asymptotics $x, t
\rightarrow \infty$ was calculated for fixed ratio $k_0 = x/2t$. The
crucial parameter for the asymptotics is the average number of particles
$xD$ in the interval $[0,x]$. If $x$ is large but $xD \ll 1$ (i.e.\ $T$
small), an electron propagates freely from $0$ to $x$, and the
correlation functions (\ref{gp1}), (\ref{gm1}) are those of free
fermions,
\begin{eqnarray} \label{gpf}
     G_f^+ (x,t) & = & \frac{e^{- \frac{{\rm i} \pi}{4}}}{2 \sqrt{\pi}}
		       \, t^{- \frac{1}{2}} e^{{\rm i} t (\mu - B)}
			  e^{\frac{{\rm i} x^2}{4t}}\ , \\ \label{gmf}
     G_f^- (x,t) & = & \frac{e^{\frac{{\rm i} \pi}{4}}}{2 \sqrt{\pi}}
	               \, e^{\frac{(\mu - B - k_0^2)}{T}}
		       \, t^{- \frac{1}{2}} e^{- {\rm i} t (\mu - B)}
			  e^{- \frac{{\rm i} x^2}{4t}}\ .
\end{eqnarray}
The true asymptotic region is characterized by a large number $xD$ of
particles in the interval $[0,x]$, specifically, $xD \gg z_c^{- 1}$,
where $z_c = (T^{3/4} e^{- k_0^2/2T})/(2 \pi^{1/4} k_0^{3/2})$. If the
latter condition is satisfied, the correlation functions decay due to
multiple scattering. The cases $B > 0$ and $B \le 0$ have to be treated
separately. For $B > 0$ there is a critical line $x = 4t \sqrt{B}$ in
the $x$-$t$ plane, which separates it into a time and a space like
regime. The asymptotics (for small $T$) in these respective regimes
is:\\[1ex]
{\bf Time like regime ($x < 4t \sqrt{B}$):}
\begin{eqnarray} \label{gpas2}
     G_{\uparrow \uparrow}^+ (x,t) & = & G_f^+ (x,t) \,
	\frac{t^{- {\rm i} \nu(z_c)} e^{- x D_\downarrow}}
	     {\sqrt{4 \pi z_c x D_\downarrow}}\ , \\ \label{gmas2}
     G_{\uparrow \uparrow}^- (x,t) & = & G_f^- (x,t) \,
	\frac{t^{{\rm i} \nu(z_c)} e^{- x D_\downarrow}}
	     {\sqrt{4 \pi z_c x D_\downarrow}}\ ,
\end{eqnarray}
where
\begin{eqnarray} \label{nuzp}
     \nu(z_c) & = & - \; \frac{2D_\downarrow k_0^{3/2} e^{- k_0^2/2T}}
                              {\pi^{1/4} T^{5/4}}\ , \\
     D_\downarrow & = & \frac{1}{2} \sqrt{\frac{T}{\pi}}
				    e^{(\mu + B)/T}\ .
\end{eqnarray}
$D_\downarrow = \partial P /\partial (\mu + B)$ is the low temperature
expression for the density of down-spin electrons.\\[1ex]
{\bf Space like regime ($x > 4t \sqrt{B}$):}
\begin{eqnarray} \label{gpas}
     G_{\uparrow \uparrow}^+ (x,t) & = & G_f^+ (x,t) \,
	t^{- {\rm i} \nu(\gamma^{-1})} e^{- x D_\downarrow}\ ,
	   \\ \label{gmas}
     G_{\uparrow \uparrow}^- (x,t) & = & G_f^- (x,t) \,
	t^{{\rm i} \nu(\gamma^{-1})} e^{- x D_\downarrow}\ ,
\end{eqnarray}
where
\begin{equation} \label{nugamma}
     \nu(\gamma^{-1}) = - \; \frac{e^{(3B + \mu - k_0^2)/T}}{2\pi}\ .
\end{equation}
For $B \le 0$ there is no distinction between time and space like
regimes. The asymptotics is given by (\ref{gpas}), (\ref{gmas}).

Equations (\ref{gpas2}), (\ref{gmas2}) and (\ref{gpas}), (\ref{gmas})
are asymptotic expansions in $t$ consisting of an exponential factor,
a power law factor and a constant factor. Note that the method employed
in \cite{GIK98} allows for a systematic calculation of the next,
subleading orders.

The leading exponential factor in (\ref{gpas2}), (\ref{gmas2}) and
(\ref{gpas}), (\ref{gmas}) has a clear physical interpretation: Because
of the specific form of the infinite repulsion in (\ref{gasham}), up-spin
electrons are only scattered by down-spin electrons. This is reflected
in the fact that the correlation length is $1/D_\downarrow$. The
exponential decay means that, due to the strong interaction, an up-spin
electron is confined by the gas of surrounding down-spin electrons.
Thus we are facing an interesting situation: Although at small distances
the electrons look like free fermions, they are confined on a
macroscopic scale set by the density $D_\downarrow$.

This work was supported by the Deutsche Forschungsgemeinschaft under
grant number Go 825/2-1 (F.G.) and by the National Science Foundation
under grant number PHY-9605226 (V.K.).

\end{multicols}


\begin{thebibliography}{10}

\bibitem{Haldane}
F.~D.~M. Haldane, Phys. Rev. Lett. {\bf 45}, 1358 (1980), and
J. Phys. C {\bf 14}, 2585 (1981).

\bibitem{BPZ84}
A.~A. Belavin, A.~M. Polyakov, and A.~B. Zamolodchikov, Nucl. Phys. B
{\bf 241}, 333 (1984).

\bibitem{YangGaudin}
C.~N. Yang, Phys. Rev. Lett. {\bf 19}, 1312 (1967),
M.~Gaudin, Phys. Lett. A {\bf 24}, 55 (1967).

\bibitem{Takahashi71b}
M.~Takahashi, Prog. Theor. Phys. {\bf 46}, 1388 (1971).

\bibitem{IzPr}
A.~G. Izergin and A.~G. Pronko, Phys. Lett. A {\bf 236}, 445 (1997),
and Nucl. Phys. B {\bf 520}, 594 (1998).

\bibitem{GIKP98}
F.~G\"ohmann, A.~G. Izergin, V.~E. Korepin, and A.~G. Pronko,
Int. J. Mod. Phys. B {\bf 12}, 2409 (1998).

\bibitem{GIK98}
F.~G\"ohmann, A.~R. Its, and V.~E. Korepin, Phys. Lett. A {\bf 249},
117 (1998).

\bibitem{Berkovich91}
A.~Berkovich, J. Phys. A {\bf 24}, 1543 (1991).

\bibitem{LiWu68}
E.~H. Lieb and F.~Y. Wu, Phys. Rev. Lett. {\bf 20}, 1445 (1968).

\bibitem{FrKo}
H.~Frahm and V.~E. Korepin, Phys. Rev. B {\bf 42}, 10553 (1990), and
Phys. Rev. B {\bf 43}, 5653 (1991).

\bibitem{Takahashi72}
M.~Takahashi, Prog. Theor. Phys. {\bf 47}, 69 (1972).

\bibitem{Takahashi74}
M.~Takahashi, Prog. Theor. Phys. {\bf 52}, 103 (1974).

\end{thebibliography}
\end{document}